\documentclass[a4paper,11pt]{article}
\usepackage{graphicx,amssymb,amstext,amsmath}
\usepackage{setspace}
\usepackage[font=small]{caption}
\begin{document}
\title{ Order-dependent structure of High Harmonic Wavefronts}
\author{E. Frumker,$^{1,2,3}$ G. G. Paulus,$^{3,4}$ H. Niikura,$^{1,5}$\\
 A. Naumov,$^{1}$ D. M. Villeneuve,$^{1}$ and P. B. Corkum$^{1}$
 \\
\normalsize{$^{1}$Joint Attosecond Science Laboratory, University of Ottawa and National Research,}\\
\normalsize{Council of Canada, 100 Sussex Drive, Ottawa, On, Canada}\\
\normalsize{$^{2}$ Max-Planck-Institut f\"{u}r Quantenoptik, Hans-Kopfermann-Strasse 1,}\\
\normalsize{ D-85748 Garching, Germany}\\
\normalsize{$^{3}$ Department of Physics, Texas A\&M University, College Station,}\\
\normalsize{Texas 77843, USA}\\
\normalsize{$^{4}$ Institute of Optics and Quantum Electronics,}\\
\normalsize{ Max-Wien-Platz 1, Jena, 07743, Germany}\\
\normalsize{$^{5}$PRESTO, Japan Science and Technology Agency,}\\
\normalsize{5 Sanbancho,Chiyodaku, Tokyo Japan 102-0075} \\
\\
}

%
\maketitle

\textbf{The physics of high harmonics has led to the generation of attosecond pulses \cite{Hentschel_Atto_Met_Nature_2001} and to trains of attosecond pulses \cite{Paul_Rabbit_Science2001}. Measurements that confirm the pulse duration are all performed in the far field. All pulse duration measurements tacitly assume that both the beam's wavefront and intensity profile are independent of frequency. However, if one or both are frequency dependent, then the retrieved pulse duration depends on the location where the measurement is made. We measure that each harmonic is very close to a Gaussian, but we also find that both the intensity profile  and the beam wavefront depend significantly on the harmonic order.
 Thus, our findings mean that the pulse duration will depend on where the pulse is observed. Measurement of spectrally resolved wavefronts along with temporal characterization at one single point in the beam would enable complete space-time reconstruction of attosecond pulses. Future attosecond science experiments need not be restricted to spatially averaged observables.}
\\



High harmonic generation is a highly coherent process in which many photons are converted into one, leaving the generating atoms in the nonlinear medium unchanged. High harmonic generation is often described with a semiclassical model in which an electron tunnel ionizes, follows a semiclassical trajectory and then recombines by emitting a single photon. In a gas of many atoms a phase matched output results in laser-like coherence of the harmonics, which has been confirmed by interfering spectrally resolved replica beams \cite{Bellini_temp_coherence_PRL1998} or different spatial regions of the same beam \cite{bartels_coherence_Science2002}. These measurements however, do not address differences in the amplitude or curvature of wavefronts between harmonics, even though we know that the wavefront depends upon the trajectory \cite{Lewenstein_phase_PRA1995,Gaarde_space_time_HHG_2quantum_path_PRA_1999,Salieres_Cohrnt_Control_HHG_PRL_1995}. Such a differences will be important for all attosecond pulse and pulse train applications. They are the subject of this paper.

To measure the spectrally and spatially resolved amplitudes and phases, we used frequency resolved wavefront characterization (or SWORD-Spectral Wavefront Optical Reconstruction by Diffraction) \cite{Frumker_sword_OL2009}. The measurement is free of approximations - it relies only on linear optics.

 Our measurements reveal that the short-trajectory harmonics have an essentially Gaussian spatial profile and that the root-mean-square deviations from a parabolic wavefront are only $\lesssim 0.2$ radians. They also reveal that different harmonics have a significantly different wavefront curvature upon emerging from the generating region.  Since we measured both amplitude and phase of each harmonic, we are able to propagate the fields and determine their spatial properties anywhere along the direction of propagation. Specifically we determine their waist position (position of the flat phase)  and size - the virtual origin of the coherent beams. We find that this virtual origin of each harmonic is at significantly different position. Particularly important, the distance between the waist positions of some harmonics can be larger than their respective Rayleigh ranges. Since the harmonics are generated in the gas jet, this observation means that they are produced with different radii-of-curvature, depending on the harmonic order. It also means that, if the harmonics are focused to measure their temporal structure, the duration of the pulse will depend on where the measurement is made.


%

\begin{figure}[htb]
\centerline{\includegraphics[width=\columnwidth]{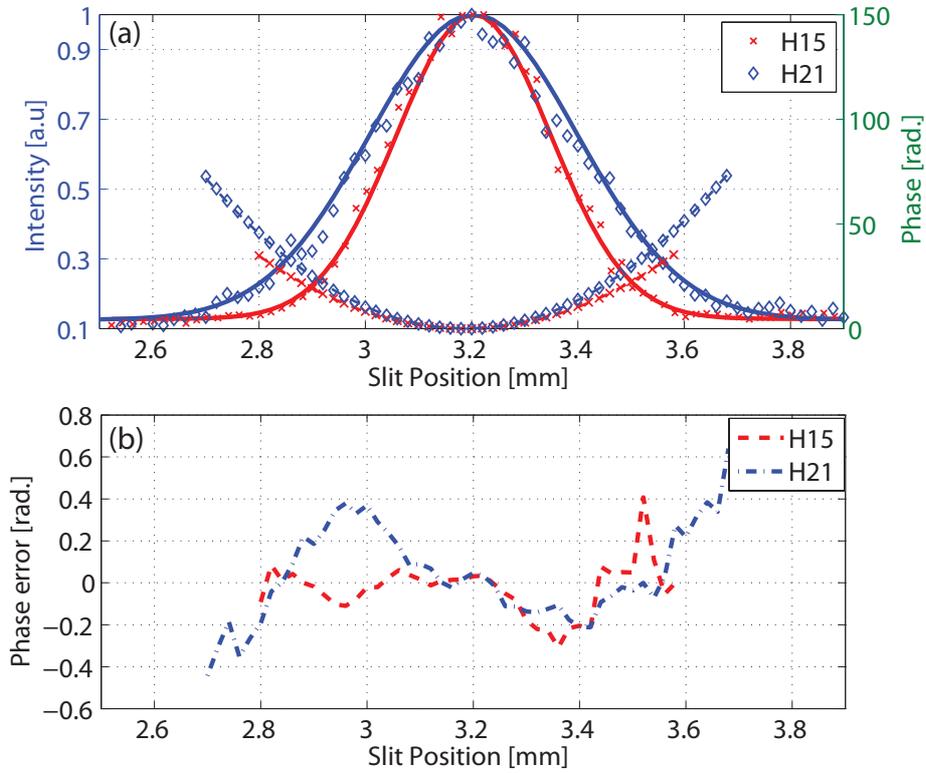}}
\caption{(a) Experimentally reconstructed spectral wavefront phase and amplitude at the slit position for different harmonics. The harmonics H15 (data points plotted as 'x') and H21 (data points plotted as '$\diamond$') are shown in red and blue respectively.  The phase data points are interpolated by the dashed line, and the measured amplitude data points are shown interpolated by solid line. The background signal arises from stray light. (b) The difference between the fitted parabolic phase and the measured phase.  }
\label{Fig_SWORD_complete_reconstruction}
\end{figure}

In our experiment we measure high harmonics produced by focusing an 800nm, 35fsec, 400$\mu$J, laser beam onto a Nitrogen gas jet. For SWORD \cite{Frumker_sword_OL2009} a small $20\mu$ wide slit is scanned across the XUV beam and the spectrally resolved diffraction pattern is recorded. Using this diffractogram, we reconstruct spectral wavefront phase and amplitudes of each harmonic.
 As the fundamental beam size in the interaction region is much smaller than the gas jet size, we can assume a uniform gas density across the beam in the generation region and therefore rotational symmetry of the generated harmonic beam.
 The further details about the experimental setup and phase reconstruction are described in the Methods section. Figure \ref{Fig_SWORD_complete_reconstruction} (a) shows examples of the reconstruction for harmonics 15 and 21.

 The intensity distribution that we measure for different harmonics in Fig. \ref{Fig_SWORD_complete_reconstruction} (a) is Gaussian, within the $\sim15\%$ error bars caused by fluctuations in the laser power and gas jet density.
 We extract the Gaussian beam parameters by fitting the measured intensity profile for each harmonic.

 The wavefront phase $\phi_N(x,z_{slit})$, where $z_{slit}$ is the slit position along the beam propagation and $x$ is the coordinate across the slit, is parabolic for all harmonic orders $N$ (Fig. \ref{Fig_SWORD_complete_reconstruction} (a)).
  Because the noise in the phase measurements is small, the error bars are not visible on the scale of Figure \ref{Fig_SWORD_complete_reconstruction} (a).
  To better estimate the phase error we subtract the fitted parabolic phase from the measured phase as shown in Figure \ref{Fig_SWORD_complete_reconstruction} (b). Across the central part of the beam, containing $95 \%$ of the beam energy, the standard deviation of measured phase from ideal parabolic phase is $0.12$rad. (better than $\lambda/50$) for the $15$th harmonic and $0.22$rad (better than $\lambda/28$) for the $21$st harmonic. Details about the fitting procedure for amplitude and phase of the wavefront are provided in the Methods section.

\begin{figure}[htb]

\centerline{\includegraphics[width=\columnwidth]{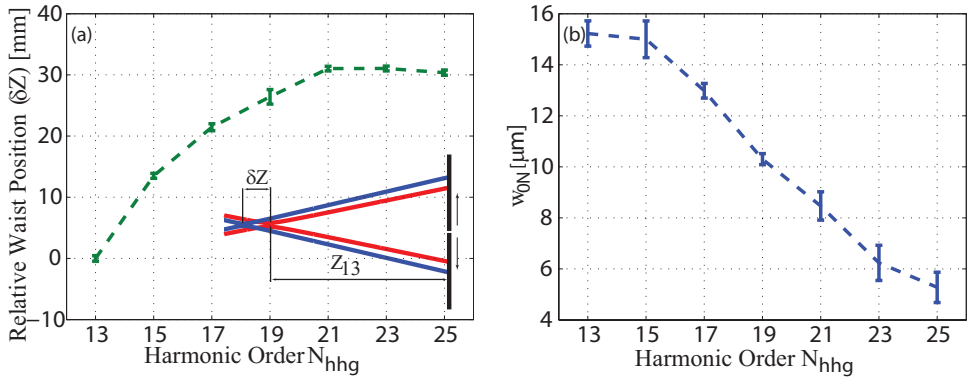}}
\caption{ (a) Relative waist position ($\delta Z$) as a function of harmonic order. The error bar is shown within $95\%$ confidence bounds. The insert illustrates that different harmonics "originate" from different positions in space. The meaning of the parameters $\delta Z$ and $Z_{13}$ is illustrated in the inset. (b) The beam waist size $w_{0N}$ as a function of harmonic order. The error bar is shown within $95\%$ confidence bounds. }
\label{Fig_waist_pos}
\end{figure}

Given the amplitude and the phase of the electric field of each harmonic at the scanning slit plane, we propagate the complex electromagnetic field backwards in time.
Gaussian beams with parabolic wavefront phase remain Gaussian as they propagate. They can be characterized by their beam waist position $Z_\textrm{N}$ (relatively to the scanning slit) and their beam waist size $w_{\textrm{0N}}$.
Figure \ref{Fig_waist_pos} (a) shows the reconstructed waist positions as a function of the harmonic order. The differences in the position $Z_\textrm{N}$ for different harmonics quantitatively show the different curvatures for each harmonic. The relative waist position $\delta Z$ (Fig. \ref{Fig_waist_pos} (a)) varies by more than their Rayleigh range  ($Z_{0N}$) which monotonically
decrease from 11.8mm for H13 to 2.7mm for H25 ($Z_{0N}=\frac{\pi w_{0N}^2}{\lambda_N}$, where $\lambda_N$ is the wavelength of $N^{th}$ harmonic). The corresponding beam (Fig. \ref{Fig_waist_pos} (b)) waists vary from $15.2 \mu m$ for H13 to $5.3 \mu m$ for H25.


\begin{figure}[ht]
\centerline{\includegraphics[width=0.74\textwidth]{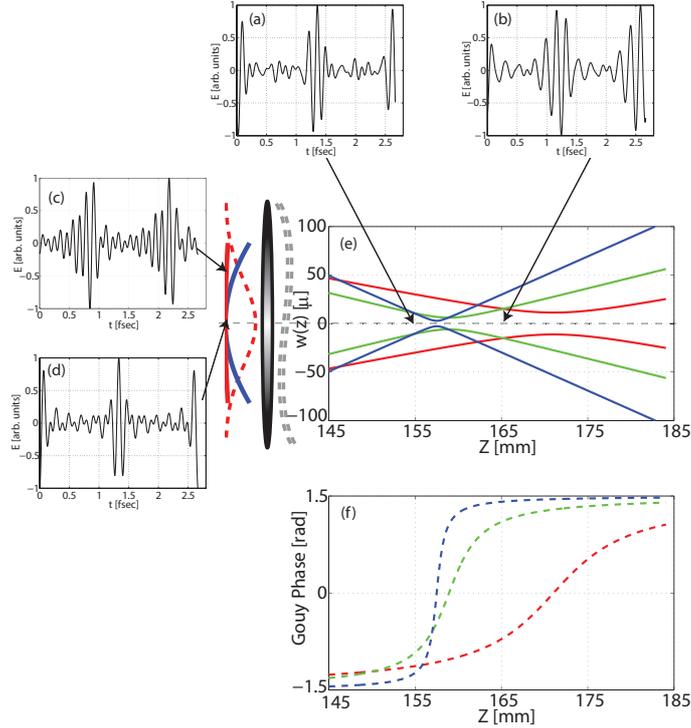}}
\caption{Spatial-temporal coupling. Panel (e) shows the selected harmonics near their focal position assuming a focusing mirror of focal length f=100mm placed a the position of the scanning slit. We see that different harmonics focus at different positions and have different divergence.
  Panel (f) shows the Gouy phase of the same harmonics around the focus. Assuming transform limited pulse at the center of the focusing mirror ($z=0$, $R=0$ - panel (d)) we can determine
  the pulses temporal profile at any position. Panels (a), (b) and (c) show respective pulses at $Z=155\textrm{mm}$, $R=0$ (panel (a)); $Z=165\textrm{mm}$, $R=0$ (panel (b)) and $Z=0$, $R=300\mu$ (panel (c)). In (a)-(d) the field strength is normalized to unity. Red, green and blue colors show the results for 13, 19 and 23 harmonic respectively.
}
\label{Fig4_focusing_problems}
\end{figure}

 The dependence of wavefront structure on harmonic order means that it is impossible to focus the attosecond pulse into one diffraction limited spot by achromatic optics (by using mirrors for example). Figure \ref{Fig4_focusing_problems} shows how the harmonics would focus if an f=100mm mirror was placed at the scanning slit position (about 250mm from the jet).

With different harmonics focusing at different longitudinal positions it is impossible to maintain a Fourier-limited attosecond pulses as the beam propagates. As soon as a particular harmonic component passes through the focus - it acquires a $\pi$ phase shift (the Gouy phase). 
 Figure \ref{Fig4_focusing_problems} (f) shows the Gouy phase across the focal region for different harmonics.
 Spectrally resolved wavefronts open the door for complete spatial-temporal reconstruction of attosecond pulses.
 Given the  measured spectral wavefront (spectral wavefront phase - $\phi_N(\vec{r},z)$, and spectral wavefront amplitude - $|E(\vec{r},z,\omega_N)|$, where $\vec{r}$ - transverse coordinate across the beam, $z$ - is the coordinate along the beam propagation) at any plane $z=z_0$ along the beam propagation, it is enough to have temporal information at \textit{any single point} $\vec{r}=\vec{r_0}$ across the beam to determine the temporal structure everywhere.

Since it is sufficient to measure the spectral phase $\varphi_N(\vec{r}_0,z_0)$ for any $\left(\vec{r}_0, z_0\right)$, we could select the point, for example, by centering a small aperture on high-harmonic beam and measuring the spectral phase of the transmitted radiation using RABBIT, FROG-CRAB \cite{Paul_Rabbit_Science2001,Mairesse_FROG_CRAB_PRA_2005} or any other suitable technique. Then, the complete spatial (3D)+ temporal (1D) field profile of the beam can be reconstructed -
 \begin{equation}\label{Eq_space_time}
 E_{hhg}(\vec{r},z,t)=\sum{|E(\vec{r},z,\omega_N)|\exp{i[\varphi_N(\vec{r}_0,z_0)+\phi_N(\vec{r},z)+N\omega_0t]}}
 \end{equation}

  where $\omega_0$ is the angular frequency of the fundamental beam and the summation is over all emitted harmonic orders.
 Note, while the spectral wavefronts $|E(\vec{r},z,\omega_N)|\exp{i[\phi_N(\vec{r},z)]}$ is measured across one plane ($z=z_0$), it can be found across any plane along the beam propagation using Fresnel-Kirchhoff diffraction formula \cite{Goodman_book_Fourier_Optics_2005} for each harmonic.

 To demonstrate this principle, we arbitrary assume that the attosecond pulses within the train are transform limited at the center of the focusing mirror (Fig. \ref{Fig4_focusing_problems} (d)) i.e. $\varphi_N=N \omega_0 t_e$, where $t_e$ is harmonic emission time. All other parameters (i.e. the wavefront structure of each harmonic) are taken from our measurement.
As we move along or across the beam, the relative spectral amplitude and phase of the harmonics changes. Consequently the temporal structure of each attosecond pulse in the pulse train depends on where the pulse is measured. Using the measured spectral wavefronts in our experiment, the temporal profiles is shown for selected points along the beam in Fig. \ref{Fig4_focusing_problems} (a)-(c). Even in cases where the atto-chirp \cite{Mairesse_atto_chirp_Science_2003,Kim_attochirp_PRA_2004} has been corrected, it can only be corrected strictly speaking at one point.

 There are at least two approaches for bringing the harmonics to a common focus and generating Fourier limited attosecond pulses and pulse trains. (1)The phase of any harmonic depends on both the intensity and the phase of generating beam. Optimizing the spatial profile (both phases and amplitudes) of the fundamental beam will allow us to minimize the chromatic wavefront variations between different harmonics.  (2)The chromatic aberrations that we measure are reproducible.  They can, in principle, be corrected with specially designed linear optics.

In conclusion, within the signal-to-noise limits of our experiment, we have obtained complete (phase and amplitude) information on wavefront profile of each harmonic. We have shown that spectrally resolved wavefronts allow the field distribution to be found anywhere.  If we know the position of the generating medium with sufficient accuracy, this includes the position at the generating medium itself.  The wave front amplitude and phase in the generating medium contains an imprint of the underlying single atom or molecule response.
Within a 3-step model \cite{Corkum_3step_PRL1993}, for a thin, low density gas, the measured wavefront phase $\phi_N(\vec{r}, z_{jet})$  is given by the sum of four contributions:
\begin{equation}\label{Eq_phase_hhg}
   \phi_N(\vec{r},z_{jet})= \phi_t(I(\vec{r},z_{jet}),\omega_N)-|\alpha_{N}|I(\vec{r},z_{jet})+\phi_r(\omega_N,z_{jet})+N\phi_{\textrm{fund}}(\vec{r},z_{jet})
\end{equation}
Where $\phi_{\textrm{fund}}(\vec{r},z_{jet})$ is the wavefront phase of the fundamental beam at the jet position. $\phi_{\textrm{fund}}(\vec{r},z_{jet})$ can be measured. $\phi_r(\omega_N,z_{jet})$ is the phase contribution of the transition moment \cite{Le_QRS_PRA2009}. $\phi_r(\omega_N,z_{jet})$ is intensity independent for a single orbital, but intensity dependent for multiple orbitals. The phase that the electron acquires in the continuum is $-|\alpha_{N}|I(\vec{r}, z_{jet})$. This phase can be calculated within the strong field approximation and its corrections \cite{Lewenstein_phase_PRA1995,Gaarde_space_time_HHG_2quantum_path_PRA_1999}. $\phi_t(I(\vec{r},z_{jet}),\omega_N)$ is the yet not fully understood tunneling phase \cite{Smirnova_multielectron_Nature2009,McFarland_N2_HOMO_1_Science2008}. It might be intensity dependent. Therefore, wavefront structure will serve as a sensitive measurement of the intensity dependence of these parameters.

Once the relative phase of each harmonic is determined, then we have shown that we know the temporal profile of an attosecond pulse anywhere in space.  No matter what the pulse temporal structure should prove to be, we have already determined that there will be significantly different temporal profile at the center and at the edges of a beam and the temporal profile must change as the beam passes through the respective foci of the harmonics. Therefore, any spatially extended measurement with attosecond pulses will be affected and any measurement of the pulse itself is also affected. The result of any experiment will be effectively averaged over the different temporal profiles at different positions in the interaction region of the focal volume.

%
%

Finally, attosecond science has been restricted to spatially averaged measurements to image orbitals \cite{Itatani_tomog_Nature2004}, tracing molecular dynamics \cite{Li_dynamics_N2O4_Science_2008}, identify and time resolve tunneling wave packets \cite{Smirnova_multielectron_Nature2009} and to follow Auger decay \cite{Drescher_Auger_Nature_2002}. Spectrally resolved wavefronts and the complete spatio-temporal characterization of attosecond pulses that they facilitate will allow us much greater experimental precision in all of these experiments.


We thank Mike Spanner, Misha Ivanov, Kyung Teac Kim and Eric Constant for stimulating discussions.
We acknowledge financial support of Canada's National Research Council, National Sciences, Engineering Research Council, the Canada Research Chair Program and MURI grant W911NF-07-1-0475. E.F acknowledges the support of Marie Curie International Outgoing Fellowship and H.N acknowledges the support of JST's Presto program.

\subsection{Methods}
The laser beam from Ti:Sapphire amplifier was spatially filtered by propagating it through hollow-core fiber in vacuum and the beam was then focused with an $f/\#=80$ lens onto the pulsed gas jet.

The gas jet in which the harmonics were generated has $250\mu m$ aperture and was operated with a backing pressure of 2.7atm. The harmonic signal was recorded using a 40mm diameter Burle imaging microchannel plate (MCP) in the imaging plane of the spectrometer and the back-side phosphor screen was imaged onto the CCD camera. Further details on experimental setup can be found in Ref. \cite{Frumker_sword_OL2009}.

The Rayleigh range of the fundamental beam ($Z_R\simeq10\textrm{mm}$) was much longer than the interaction region of the jet ($d\simeq0.4\textrm{mm}$) where harmonics were generated. The intensity of the fundamental beam was kept moderate to prevent excessive ionization and plasma creation. This configuration is particularly useful when we aim to study single atom/molecular response and has been widely used in attosecond science \cite{Itatani_tomog_Nature2004}.

The high harmonic radiation was diffracted through a horizontal scanning slit positioned in front of the XUV spectrometer \cite{Frumker_sword_OL2009}. The spectrometer is built with variable groove spacing flat field Hitachi imaging grating (model 001-0266). The system resolves the spectrum in horizontal direction, and allows essentially free-space field propagation in vertical direction. Two-dimensional images are taken for each position of the scanning slit using imaging multichannel plate (MCP) and cooled CCD camera.

The relative vertical position of the diffraction pattern's centroid is proportional to the wavefront slope of the sampled wavefront slice. We reconstruct the wavefront phase in Fig. \ref{Fig_SWORD_complete_reconstruction} (a) using the following relation: $\delta_j=\delta{z}\cdot \frac{y_j-z_j}{d}$, where $y_j$ - is the relative position of the centroid of the diffracted pattern, $z_j$ - relative slit position, $d$ - is the distance between the scanning slit and the imaging plane, $\delta{z}$ - is the scanning slit iteration step, and $\delta_j$ - is the optical path  difference across the slit.
Then we determine the sampled wavefront phase profile - $\phi^{i}_{Nexp}$ for the harmonic wavelength - $\lambda_N$ across the scanning direction using $\phi^{i}_{Nexp}= \frac{2\pi}{\lambda_N}\sum_{j=0}^{i}\delta_j$.
We determine the wavefront amplitude $I^{i}_{exp}$ at each sampling point by integrating intensity of the corresponding diffraction pattern. Further details of SWORD are discussed in Ref. \cite{Frumker_sword_OL2009}.
%

The experimental wavefront phase $\{\phi^{i}_{Nexp}\}$  sampled at positions $\{x^{i}_{exp}\}$ was fit with parabolic function $\phi_N(x)=a_N (x-x_0)^2$ for each harmonic. The Nonlinear Least Square method was used for the fit. To minimize the noise contribution, the phase is fit within the range where measured intensity is larger than $5\%$ relative to the peak intensity. The wavefront intensity distribution $\{I^{i}_{exp}\}$ measured at the scanning slit was fit to the Gaussian profile $I(x)={I_0}\exp[-2(x-x_0)^2/{w_{sN}^2}]+I_{b}$ , where $I_0$ is maximum intensity, $w_{sN}$ - Gaussian width at the slit position, and $I_b$ - background signal.  The measured parabola curvature - $a_N$, and the Gaussian width - $w_{sN}$ for each harmonic (H13-H25) were used to retrieve the virtual waist position $Z_N$ and waist size $w_{0N}$. A sensitivity analysis showed that the $Z_N$ is highly sensitive to $a_N$ and depends weekly on $w_{sN}$ (ex.: $10\%$ change in the Gaussian width causes only $\sim0.1\%$ change in $Z_N$). We use the confidence bound of $95\%$ in determination of the error bars for the relative waist position $\delta Z$ with respect to the scanning slit as shown in Fig \ref{Fig_waist_pos}.

 We have measured short trajectory harmonics. The divergence of the short trajectories mainly depends on the position of the jet relative to focal position of the fundamental beam. For all jet positions we find that the virtual origin of the harmonics differs by more than their confocal parameter across the spectrum.

\bibliographystyle{unsrt}
\bibliography{D:/NRC_work_since_December2007/PostDoc_Papers/Orientation_nature/Orientation_Science/Atto_references}

%
%
%

\end{document}